\documentclass[prd,twocolumn,showpacs,superscriptaddress,eqsecnum]{revtex4}
\usepackage{graphicx}

\newcommand{\mev}{\rm{MeV}}

\newcommand{\cm}{\rm{cm}}

\newcommand{\g}{\rm{g}}

\newcommand{\nua}{{\nu_\alpha}}
\newcommand{\phij}{\phi_j}
\newcommand{\phie}{\phi_{\nu_e}}
\newcommand{\phimu}{\phi_{\nu_\mu}}
\newcommand{\phitau}{\phi_{\nu_\tau}}

\newcommand{\one}{ 1\!\!1}
\newcommand{\nue}{\nu_e}
\newcommand{\numu}{\nu_\mu}
\newcommand{\nutau}{\nu_\tau}

\begin{document}

\title{Measuring Flavor Ratios of High-Energy Astrophysical Neutrinos}

\author{John F. Beacom}
\email{beacom@fnal.gov}
\affiliation{NASA/Fermilab Astrophysics Center, Fermi National Accelerator
Laboratory, Batavia, Illinois 60510-0500}

\author{Nicole F. Bell}
\email{nfb@fnal.gov}
\affiliation{NASA/Fermilab Astrophysics Center, Fermi National Accelerator
Laboratory, Batavia, Illinois 60510-0500}
\affiliation{Kavli Institute for Theoretical Physics, University of
California, Santa Barbara, California 93106}

\author{Dan Hooper}
\email{hooper@pheno.physics.wisc.edu}
\affiliation{Department of Physics, University of Wisconsin,
Madison, Wisconsin 53706}

\author{Sandip Pakvasa}
\email{pakvasa@phys.hawaii.edu}
\affiliation{Department of Physics and Astronomy, University of Hawaii,
Honolulu, Hawaii 96822}
\affiliation{Kavli Institute for Theoretical Physics, University of
California, Santa Barbara, California 93106}

\author{Thomas J. Weiler}
\email{tom.weiler@vanderbilt.edu}
\affiliation{Department of Physics and Astronomy, Vanderbilt University,
Nashville, Tennessee 37235}
\affiliation{Kavli Institute for Theoretical Physics, University of
California, Santa Barbara, California 93106}

\date{July 1, 2003}

\begin{abstract}
We discuss the prospects for next generation neutrino telescopes, such
as IceCube, to measure the flavor ratios of high-energy astrophysical
neutrinos.  The expected flavor ratios at the sources are
$\phi_{\nu_e}:\phi_{\nu_{\mu}}:\phi_{\nu_{\tau}} = 1:2:0$, and
neutrino oscillations quickly transform these to $1:1:1$.  The flavor
ratios can be deduced from the relative rates of showers ($\nu_e$
charged-current, most $\nu_\tau$ charged-current, and all flavors
neutral-current), muon tracks ($\nu_\mu$ charged-current only), and
tau lepton lollipops and double-bangs ($\nu_\tau$ charged-current
only).  The peak sensitivities for these interactions are at different
neutrino energies, but the flavor ratios can be reliably connected by
a reasonable measurement of the spectrum shape.  Measurement of the
astrophysical neutrino flavor ratios tests the assumed production
mechanism and also provides a very long baseline test of a number of
exotic scenarios, including neutrino decay, CPT violation, and
small-$\delta m^2$ oscillations to sterile neutrinos.
\end{abstract}

\pacs{95.85.Ry, 96.40.Tv, 13.35.Hb, 14.60.Pq
\hspace{0.5cm} FERMILAB-Pub-03/180-A, MADPH-03-1336}


\maketitle


\section{Introduction}

A new generation of detectors proposed or already under construction
will have the sensitivity to open a new window on the universe in the
form of high-energy neutrino astronomy.  Besides probing distant and
mysterious astrophysical sources, these data will offer unprecedented
sensitivity for testing fundamental neutrino properties.

Under-ice and under-water optical \v{C}erenkov detectors are sensitive
at roughly TeV-PeV neutrino energies.  The AMANDA
detector~\cite{AMANDA} at the South Pole has already been taking data
for several years, as has the smaller Lake Baikal
detector~\cite{Baikal}.  First results from the AMANDA-B10 phase have
been reported~\cite{AMANDA,AMANDAB10}, and results from several years
of running in the larger AMANDA-II configuration~\cite{AMANDAII} are
eagerly awaited.  Construction of the much larger (km$^3$) IceCube
detector~\cite{IceCube} at the South Pole is scheduled to begin later
this year.  The ANTARES~\cite{ANTARES} and NESTOR~\cite{NESTOR}
collaborations, drawing on the lessons learned by the DUMAND
project~\cite{DUMAND}, are currently deploying their detectors in the
Mediterranean Sea.  While smaller than IceCube, these detectors will
have lower thresholds due to denser instrumentation.  Another
Mediterranean detector, NEMO~\cite{NEMO}, has been proposed.  These
detectors operate under water or ice to shield the atmospheric muon
flux; since they primarily look for upgoing neutrino-induced events,
the detectors in the southern and northern hemispheres offer
complementary views of the sky.

Detectors making use of the Earth's atmosphere as a target volume are
sensitive at roughly EeV-ZeV neutrino energies.  Fly's
Eye~\cite{FlysEye} and AGASA~\cite{AGASA}, designed to detect
atmospheric showers induced by cosmic ray protons, are also sensitive
to ultra-high energy neutrino primaries if they consider penetrating
horizontal showers, i.e., events that initiate only at great slant
depth in the atmosphere.  The bounds on ultra-high energy neutrino
fluxes are reviewed in Ref.~\cite{UHEbounds}.  The HiRes detector (an
upgrade of Fly's Eye) is running~\cite{HiRes}, and the Pierre Auger
detector~\cite{Auger}, scheduled to be completed in 2005, is already
taking data with their engineering array.  Both should significantly
improve the sensitivity to ultra-high energy astrophysical neutrinos.
Orbiting detectors will soon provide a new window on neutrinos as
well.  The EUSO experiment~\cite{EUSO} is scheduled for
deployment on the International Space Station in early 2009, and it
may evolve into a larger satellite mission named OWL~\cite{OWL}.

Radio \v{C}erenkov detectors are sensitive at roughly PeV-ZeV neutrino
energies.  As pointed out long ago by Askaryan~\cite{Askaryan}, a
\v{C}erenkov signal is proportional to the square of the net charge of
the shower within a wavelength.  In turn, the net charge is roughly
proportional to the shower energy.  Thus, the long-wavelength radio
signal rises as the neutrino energy squared, instead of just the
neutrino energy, as for other techniques.  The RICE antennas,
co-deployed with AMANDA, have recorded first results~\cite{RICE}.
Recently, NASA approved a very interesting experiment,
ANITA~\cite{ANITA}, which will carry radio antennas on a balloon above
the South Pole in an attempt to detect the long-wavelength tail of
sub-ice showers from Earth-skimming neutrinos~\cite{earthskimming}.
The GLUE experiment has reported limits based on the non-observation
of a radio signal from neutrino interactions in the surface of the
Moon~\cite{GLUE}.  The proposed SALSA detector~\cite{SALSA} could use
large salt domes as a radio \v{C}erenkov medium.

Although the goals of these experiments are common, the detection
strategies and systematic issues are not.  Each will infer the
direction of arriving events, thereby enabling point-source neutrino
astronomy.  Each will measure the neutrino flux over some energy
range.  One purpose of this paper is to ask how well the primary
neutrino spectrum can be reconstructed from the observed spectral
information.  The primary spectrum reveals the dynamics of the cosmic
engine.  Reviews of high-energy neutrino astronomy are listed in
Ref.~\cite{reviews}.

Besides energy and direction, an additional piece of information is
carried by arriving neutrinos: flavor.  Neutrinos are known to come in
three flavors, electron-, muon-, and tau-type.  The bulk of this paper
focuses on what we can learn from flavor identification of the
incoming neutrinos, and whether flavor-tagging is feasible with
proposed experiments.  We consider the IceCube experiment in detail
due to its large effective volume and ability to observe showers, muon
tracks and events unique to tau neutrinos.  We emphasize that other
experiments will also have some capabilities to discriminate among
neutrino flavors.

In the next section, we look at theoretical motivations for flavor
discrimination.  In the sections after, we look at inferences of
the incident neutrino spectrum, and the neutrino flavors
from various event signatures.


\section{Why Neutrino Flavor Identification Is Interesting}

Neutrinos from astrophysical sources are expected to arise dominantly
from the decays of charged pions (and kaons) and their muon daughters,
which results in initial flavor ratios, $\phie:\phimu:\phitau$, of
nearly 1:2:0.  The fluxes of each mass eigenstate are then given by
$\phij=\sum_\alpha\,\phi^{\rm source}_\nua |U_{\alpha j}|^2$.  The
$U_{\alpha j}$ are elements of the neutrino mass-to-flavor mixing
matrix, defined by $|\nua\rangle = \sum_j U_{\alpha j} |\nu_j\rangle$.
The propagating mass eigenstates acquire relative phases, giving rise
to flavor oscillations.  However, these relative phases are lost,
since $\delta m^2 \times L/E \gg 1$, and hence uncertainties in the
distance $L$ and the energy $E$ will wash out the relative phases.
Thus the neutrinos arriving at Earth are an incoherent mixture of mass
eigenstates with the proportions given above.

For three neutrino species, as we assume throughout, there is now
strong evidence from atmospheric and reactor neutrino data suggesting
that $\numu$ and $\nutau$ are maximally mixed and $U_{e3}$ is nearly
zero.  This twin happenstance ($\theta_{\rm{atm}}=45^{\circ}$ and
$U_{e3}=0$) leads to two remarkable conclusions.  The first is that
each mass eigenstates contains an equal fraction of $\nu_{\mu}$ and
$\nu_{\tau}$.  The second is that in the mass eigenstate basis, the
neutrinos are produced in the ratios $1:1:1$, independent of the solar
mixing angle, and thus arrive at Earth as an incoherent mixture of
mass eigenstates with these same ratios.  This implies democracy in
the detected flavor ratios as well, since since $U\one U^\dag=\one$ in
any basis.

So there is a fairly robust prediction of $1:1:1$ flavor ratios for
measurements of astrophysical neutrinos~\cite{LP,athar1}.  The first
task of flavor measurement is to check this prediction.  Could the
flavor ratios differ from $1:1:1$?  The astrophysics could be
different than is outlined here.  For example, if the charged pion
decays promptly but the daughter muon loses energy before decaying,
then at high energy the flux may approximate $\phie:\phimu:\phitau
\sim 0:1:0$~\cite{cooling}.  This leads to mass eigenstate ratios of
$\sim 1:2:3$, and measured flavor ratios of $\sim 1:2:2$.

Alternatively, the particle physics could be different than assumed
here (three stable neutrinos).  For example, the heavier mass
eigenstates could decay en route to Earth.  This leads to markedly
different detected flavor ratios, as extreme as either $6:1:1$ or
$0:1:1$ for the normal and inverted hierarchies,
respectively~\cite{decay}.  Also, Barenboim and Quigg have pointed out
that CPT violation in neutrino mixing could also lead to anomalous
flavor ratios~\cite{Barenboim}.  Finally, neutrinos could be
pseudo-Dirac states, in which case the three active neutrinos have
sterile partners, with which they are maximally mixed with very tiny
$\delta m^2$ splittings; these oscillations might only be effective
over cosmological distances~\cite{pDirac}.

In all cases investigated, the $\nu_\mu - \nu_\tau$ symmetry ensures
that $\numu$ and $\nutau$ arrive at Earth in equal numbers.  Given how
robust that prediction is, and that $\nu_\tau$ is the most difficult
flavor to identify in IceCube, we will mostly focus on the
$\phie:\phimu$ ratio.  We will show that this can be determined by
measuring the rates of shower and track events in IceCube.
Additionally, we discuss the identification of $\nu_\tau$; though the
expected yields are very small, the signals are very distinctive, and
the detection of even a single event would confirm the presence of the
$\nu_\tau$ flux.


\section{Basics First: Measuring the Neutrino Spectrum with Muons}

For under-ice or under-water detectors, muon events provide the most
useful signal from which to infer the neutrino spectrum.  This is
especially true at lower energies (100 GeV-several TeV) due to the
higher energy threshold for showers.  In this section we explore
reconstruction of the neutrino spectrum from observed muon events in
an IceCube-type detector.  There are challenges in doing this,
however.

First, muons can be created in interactions far from the detector, and
lose a considerable fraction of their energy before being measured.
This problem can be circumvented, however, by considering only muon
events with a contained vertex, in which the muon track begins within
the detector volume.  At energies near or below the TeV scale, many of
the observed muons will have contained vertices.  At higher energies,
when the range of muons is considerably longer, fewer of the resulting
events will have this feature.

Second, separating the atmospheric neutrino background from any
astrophysical neutrino signal can be difficult.  The atmospheric
neutrino spectrum is well-modeled, and so in principle can be
subtracted from the data.  At energies at or above about 100 TeV, the
astrophysical neutrino flux is likely to be above the more steeply
falling atmospheric neutrino flux.  At more modest energies, the
angular and temporal resolution of a neutrino telescope
will be needed to effectively remove backgrounds.
The atmospheric neutrino flux at 1 TeV (dominated by $\nu_\mu +
\bar{\nu}_\mu$) is $\sim 10^{-8}$ GeV cm$^{-2}$ s$^{-1}$~\cite{agrawal}
in a $(1 {\rm\ deg})^2$ bin, where the bin size was chosen to
reflect the angular resolution for these events.  Especially for the
lower flux we consider below, identification of astrophysical sources
may require temporal information as well.
For example, gamma-ray bursts
typically have durations on the order of seconds.  Taking the known
catalogs of gamma-ray bursts as a guide, about $10$ events per square
kilometer per year are expected.  Using timing and directional
information, these events are essentially background
free~\cite{guetta}.
Tau neutrino detection for gamma-ray bursts
has been studied in Ref.~\cite{gupta}.
A similar technique could be used for blazars or
other transient sources.


\subsection{Muon Tracks}

\begin{figure}[t]
\includegraphics[height=3.25in,angle=90]{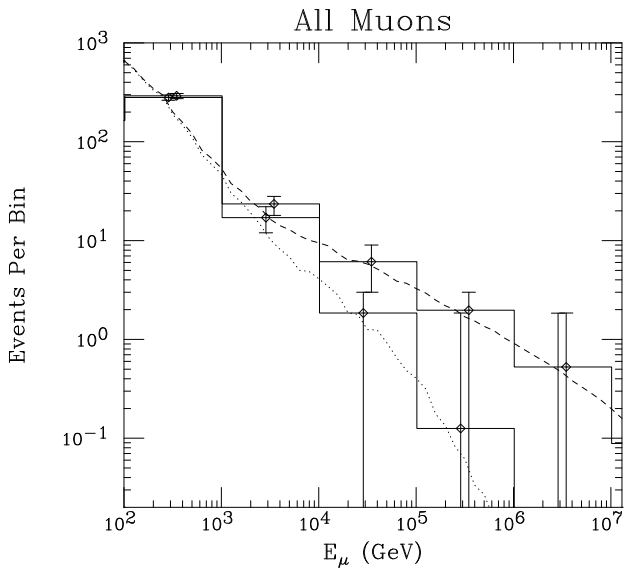}
\caption{\label{fig:one} The distribution of observed muon energies
for a neutrino spectrum of
$E^2_{\nu_{\mu}} dN_{\nu_{\mu}}/dE_{\nu_{\mu}} = 
10^{-7} \rm{\ GeV} \rm{\ cm}^{-2} \rm{\ s}^{-1}$ for one year.
The dashed line is for a flux at a horizontal zenith angle, and the
dotted line for an upgoing flux (through the Earth).  Note that the
muon energy at production may be considerably larger than that
observed due to the long muon range.  The error bars are slightly
offset for clarity.}
\end{figure}

After a high energy muon is produced, it undergoes continuous energy
loss as it propagates, given by
\begin{equation}
\frac{dE}{dX}=-\alpha - \beta E \ ,
\end{equation}
where $\alpha=2.0~\mev~\cm^2/\g$ and $\beta=4.2 \times
10^{-6}~\cm^2/\g$~\cite{propagation}. The muon range is then
\begin{equation}  
R_\mu = \frac{1}{\beta} \ln \left[ 
\frac{\alpha + \beta E_\mu}{\alpha + \beta E^{\rm thr}_\mu} \right]\\,
\label{murange}
\end{equation}
where $E^{\rm thr}_\mu$ is the minimum muon energy triggering the
detector.  Typically, $E_{\mu}^{\rm{thr}} \sim 50-100$ GeV for deep
ice or water detectors.  Above an energy of a TeV, the muon range
rises logarithmically as $\sim \ln (E_\mu/E^{\rm thr}_\mu)$ times
2.4~km water equivalent.  Since this typically exceeds the size of the
detector, the muon energy cannot be measured by the muon range.
Further, since the muons are always fully relativistic, the
\v{C}erenkov angle and intensity are constant and thus cannot be used
to infer the muon energy.  However, the muon energy inside the
detector can be inferred by the rate of energy deposition in the form
of showers from catastrophic bremsstrahlung~\cite{IceCube,CB}.  The
muon range is substantially less than the muon decay length; for an
illustration of the length scales for neutrino interactions, mu and
tau range, and mu and tau decay, see Fig.~1 of Ref.~\cite{regen2}.

\begin{figure}[t]
\includegraphics[height=3.25in,angle=90]{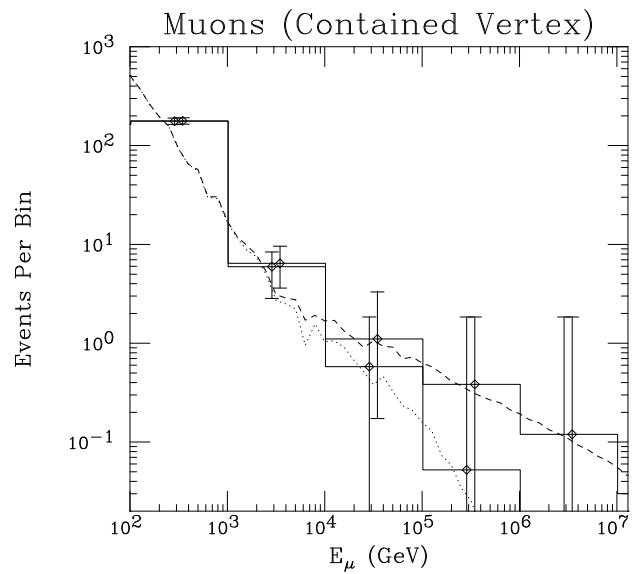}
\caption{\label{fig:two} The distribution of observed 
{\it contained vertex} muon energies for a neutrino spectrum of
$E^2_{\nu_{\mu}} dN_{\nu_{\mu}}/dE_{\nu_{\mu}} =
10^{-7} \rm{\ GeV} \rm{\ cm}^{-2} \rm{\ s}^{-1}$ for one year.
The dashed line is for a flux at a horizontal zenith angle, and the
dotted line for an upgoing flux (through the Earth).  The error bars
are slightly offset for clarity.}
\end{figure}

The energy of the muon faithfully represents the neutrino energy since
the charged-current differential cross section is strongly peaked at
$y = 1 - E_\mu/E_\nu \simeq 0$, and $\langle y \rangle \simeq
0.2$~\cite{cross}.  The kinematical angle of the muon relative to the
neutrino direction is about $1^\circ/\sqrt{E_\nu/{\rm 1\ TeV}}$, and
the reconstruction error on the muon direction is on the order of
$1^\circ$.

The probability of detecting a muon neutrino traveling through the
detector via a charged-current interaction is then given by
\begin{equation}
P_{\nu_{\mu} \rightarrow \mu}
\simeq \rho N_A \sigma R_{\mu}\,,
\label{muprob}
\end{equation}
where $\rho$ is the target nucleon density, $N_A$ is Avogadro's
number, and $\sigma$ is the neutrino-nucleon total cross
section~\cite{cross}.


\subsection{Spectral Results}

\begin{figure}[t]
\includegraphics[height=3.25in,angle=90]{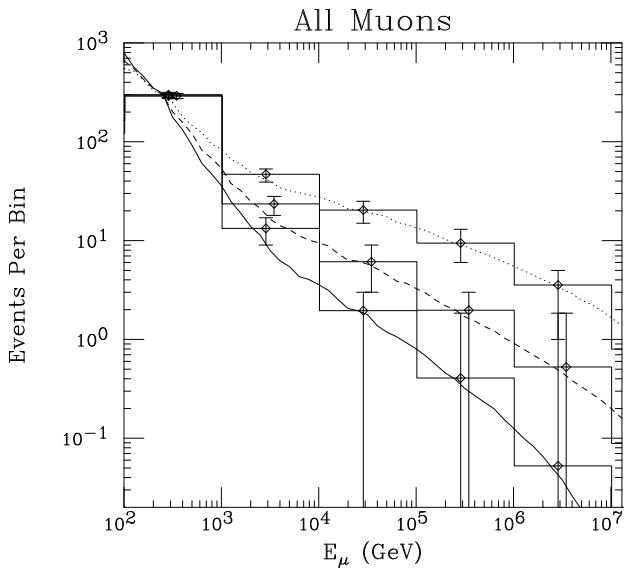}
\caption{\label{fig:three} The distribution of observed muon energies
for a neutrino spectrum of
$E^2_{\nu_{\mu}} dN_{\nu_{\mu}}/dE_{\nu_{\mu}} =
10^{-7} \rm{\ GeV} \rm{\ cm}^{-2} \rm{\ s}^{-1}$ for one year
(dashed), as in Figs.~\ref{fig:one} and \ref{fig:two}, compared
against spectra proportional to $E^{-2.2}$ (solid) and $E^{-1.8}$
(dotted), normalized to the same number of events in the first energy
bin.  All rates are for a horizontal zenith angle source.  Note that
the muon energy at production may be considerably larger than that
observed.  The error bars are slightly offset for clarity.}
\end{figure}

In Figs.~\ref{fig:one} and~\ref{fig:two}, we show the distribution of
observed muon energies for a muon neutrino spectrum of
$E^2_{\nu_{\mu}} dN_{\nu_{\mu}}/dE_{\nu_{\mu}} =
10^{-7} \rm{\ GeV} \rm{\ cm}^{-2} \rm{\ s}^{-1}$ for one year 
(or $E^2_{\nu_{\mu}} dN_{\nu_{\mu}}/dE_{\nu_{\mu}} =
10^{-8} \rm{\ GeV} \rm{\ cm}^{-2} \rm{\ s}^{-1}$ for ten years).
The dashed lines correspond to the flux at a horizontal zenith angle,
and the dotted lines to a upgoing flux (through the Earth).  Detection
prospects for horizontal neutrinos are enhanced from long distances of
ice in which muons can be produced. Upgoing neutrinos, although also
with this advantage, can be absorbed in the Earth, degrading their
event rate.  A distribution of neutrino sources over the sky produces
a spectrum in between these two extreme curves.  Fig.~\ref{fig:one}
shows the spectrum of all observed muons, while Fig.~\ref{fig:two}
shows only muon events with a contained vertex. The events are binned
by energy decade, with 68\% confidence levels shown.  The
dashed and dotted lines,
shown for comparison, are the results of our Monte Carlo with much
greater statistics. Note that the sizes of the energy bins was
selected for statistical purposes. The energy resolution of neutrino
telescopes is considerably more precise at these energies.

\begin{figure}[t]
\includegraphics[height=3.25in,angle=90]{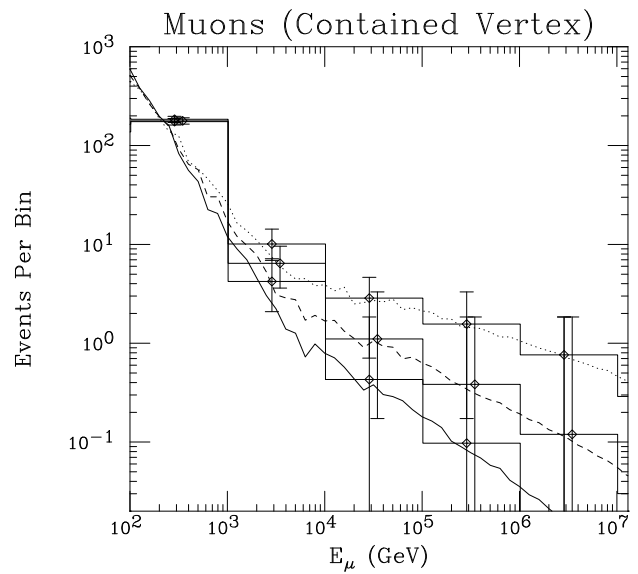}
\caption{\label{fig:four} The distribution of observed 
{\it contained vertex} muon energies for a neutrino spectrum of 
$E^2_{\nu_{\mu}} dN_{\nu_{\mu}}/dE_{\nu_{\mu}} = 
10^{-7} \rm{\ GeV} \rm{\ cm}^{-2} \rm{\ s}^{-1}$ for one year
(dashed), as in Figs.~\ref{fig:one} and \ref{fig:two}, compared
against spectra proportional to $E^{-2.2}$ (solid) and $E^{-1.8}$
(dotted), normalized to the same number of events in the first energy
bin.  All rates are for a horizontal zenith angle source.  The error
bars are slightly offset for clarity.}
\end{figure}

In Figs.~\ref{fig:three} and \ref{fig:four}, we compare the observable
horizontal muon spectra resulting from neutrino spectra proportional
to $E^{-2}$ (dashed), $E^{-2.2}$ (solid), and $E^{-1.8}$ (dotted),
normalized to the same number of events in the first energy bin.  By
comparing the numbers of events in adjacent bins, Fig.~\ref{fig:three}
demonstrates that for a single power-law flux (and our choice of
normalization), the spectral slope can be determined to approximately
ten percent up to tens or perhaps hundreds of TeV.
Fig.~\ref{fig:four} demonstrates the ability to make such as
measurement with only muons with contained vertices.  Comparing Figs.\
\ref{fig:three} and \ref{fig:four} makes it clear that limiting the
data to muons with contained vertices alone weakens the ability to
resolve similar spectral slopes, especially at higher energies.  This
is because of the statistical limitations.  However, the fact that
muons with contained vertices yield the total muon energy, whereas
throughgoing muons offer only lower limits to the energy, has its
virtue.  The contained-muons measurement is more useful for the case
of a strongly broken power-law, such as is predicted in the case of
gamma-ray bursts.  Typically, the gamma-ray burst neutrino spectrum
is expected to follow a $E^{-1}$ spectrum up to some break energy
on the order of hundreds of TeV. Above this energy, it steepens to a
spectrum proportional to $E^{-2}$. For such a spectrum, the break
could be observed with the contained vertex muons measurement, and the
slopes could be then measured more precisely using all observed muons.

We note that the flux we employ for illustration, 
$E^2_{\nu_{\mu}} dN_{\nu_{\mu}}/dE_{\nu_{\mu}} = 
10^{-7} \rm{\ GeV} \rm{\ cm}^{-2} \rm{\ s}^{-1}$,
is on the order of the Waxman-Bahcall bound~\cite{wbbound}.  The
Waxman-Bahcall bound pertains to the diffuse neutrino flux from
sources optically thin to protons, normalized to the measured
cosmic-ray flux at $\sim10^{18}$~eV.  As emphasized by Mannheim,
Protheroe and Rachen \cite{mprbound}, it does not apply for optically
thick, or ``hidden'' sources~\cite{hidden}, for galactic sources,
or for sources not emitting protons at energies $\sim 10^{18}$~eV.
Microquasars and supernovae remnants provide examples of both of the
latter categories. The current experimental limit on the high-energy
neutrino flux, from the AMANDA experiment, is more than an order of
magnitude larger than the Waxman-Bahcall
bound~\cite{AMANDA,AMANDAB10}.  IceCube is expected to reach well
below the Waxman-Bahcall bound~\cite{IceCube}.

Nature's flux could be larger than we assume here.  It could also be
smaller.  If it is smaller, then integration times larger than the one
year we assume here are needed to compensate.  Furthermore, larger
detectors, such as an extension of IceCube, are likely to be
constructed in the future, making more conservative choices of the
neutrino flux easier to study.  To reduce backgrounds, only those
events associated with known sources should be considered.  In that
case, the remaining flux of neutrinos which could be used for such a
study may be significantly reduced.  Given this consideration,
perhaps a choice of
$E^2_{\nu_{\mu}} dN_{\nu_{\mu}}/dE_{\nu_{\mu}} =
10^{-8} \rm{\ GeV} \rm{\ cm}^{-2} \rm{\ s}^{-1}$ over ten years,
which we also discuss, could be considered a more realistic choice.


\section{Flavor Identification}

\begin{figure}[t]
\includegraphics[height=3.25in,angle=90]{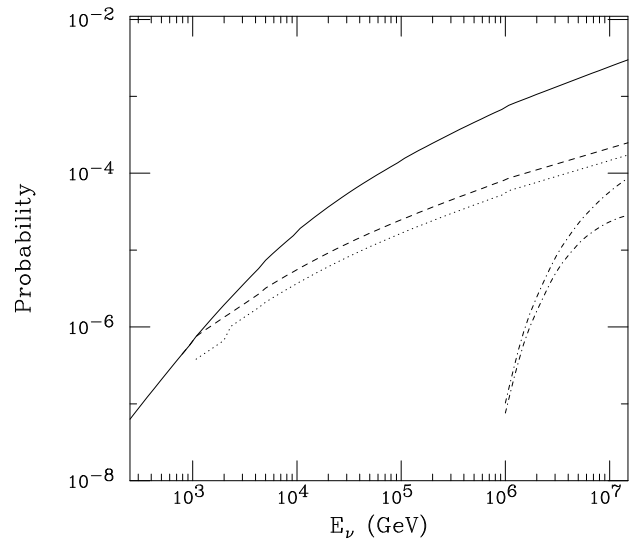}
\caption{\label{fig:probs} Probabilities of detecting different
flavors of neutrinos in IceCube versus neutrino energy, described in
detail in the text.  The upper solid line is the probability of a
horizontal $\nu_\mu$ creating a detectable muon track, and the dashed
line is for downgoing $\nu_\mu$.  The dotted line is the probability
for $\nu_e$ to create a detectable shower (above 1 TeV), considering
both charged-current and neutral-current interactions; the kink occurs
when the neutral-current showers come above threshold.  The dot-dashed
lines are the probabilities for $\nu_\tau$ to make lollipop events
(upper) and double-bang events (lower).}
\end{figure}

Although the ratios of neutrino flavors are not directly measurable,
they can be inferred from complementary classes of events in neutrino
telescopes.  In this section, we restrict our attention to IceCube,
which will be capable of identifying showers from both charged- and
neutral-current events, muon tracks, and certain tau neutrino events.
Muon tracks have been discussed above.  The probabilities for
detecting the different neutrino flavors are illustrated in
Fig.~\ref{fig:probs}, and will be discussed in detail now.

Electron and muon neutrinos above about 100 TeV are absorbed in Earth
by their charged-current interactions.  Tau neutrinos also interact,
but regenerate $\nu_\tau$ by the prompt decays of tau
leptons~\cite{regen1}; these decays also produce a secondary flux of
$\nu_e$ and $\nu_\mu$~\cite{regen2,regen3}.

\subsection{Showers}

All neutrino flavors undergo an identical neutral-current interaction
producing a hadronic shower and nothing else.  The shower energy
typically underestimates the neutrino energy by a factor ranging from
$\sim 3$ around a TeV, to a factor of $\sim 4$ at an EeV~\cite{cross}.

A charged-current interaction of $\nu_e$ produces an electron that
immediately creates an electromagnetic shower.  If the electromagnetic
shower is measured with the hadronic one, the total shower energy is
the incident $\nue$ energy.  In principle, electromagnetic and
hadronic showers are distinguishable by their respective muon content,
absent for electromagnetic and present for hadronic showers.  We will
not assume that these can be distinguished, as it is expected to be
very difficult.

In a $\nu_\mu$ charged-current interaction, the muon track always
emerges from the shower, because of the long muon range, so these
events don't contribute to the shower rate.

The $\nu_\tau$ charged-current interaction produces a hadronic shower
and a tau track.  The tau decay pathlength is $\gamma c \tau \sim
50\,(E_\tau/{\rm\ PeV})$ m.  Below a few PeV, the tau track is too
short to be separated from the shower, and so these events will
contribute to the shower rate.  At higher energies, the tau track will
extend beyond the initial shower, and then the tau will decay to
produce a second shower.  This creates identifiable double-bang and 
lollipop events, discussed below.

Showers are seen by the detector as photoelectrons distributed over a
$\sim 100$ m radius sphere for a TeV shower ($\sim 300$ m radius for
showers with PeV energies).  The shower must at least be partially
contained within the detector volume in order to be detected.  Since
shower sizes are relatively small compared to muon ranges, the
effective volume for these events is substantially less than for
charged-current $\nu_\mu$ interactions.  Also, the energy threshold
for showers is generally larger than for muon tracks.

The probability of detecting a neutrino by a shower produced by a
neutral-current interaction is given by
\begin{equation}
P_{\nu \rightarrow \rm{shower}} 
\simeq \rho N_A L 
\int^{1}_{E_{\rm{sh}}^{\rm{thr}}/E_{\nu}} \frac{d \sigma}{d y} dy
\end{equation}
where $\sigma$ is the neutrino-nucleon cross section~\cite{cross}, $y$
is the energy fraction transferred from the initial neutrino to the
hadronic shower, and $L$ is the length of the detector.  For
charged-current electron neutrino interactions, however, the
additional electromagnetic shower means that all of the neutrino
energy goes into the shower, and
\begin{equation}
P_{\nu \rightarrow \rm{shower}} \simeq \rho N_A \sigma L\,.
\label{showerprob}
\end{equation}
A similar treatment is used for charged-current tau neutrino
interactions which do not produce a double-bang or lollipop event (see
below).  In real experiments, the shower threshold is not a step
function as we adopt here.
IceCube will have shower energy resolution of about $\pm 0.1$ on a
$\log_{10}$ scale and will be able to reconstruct the neutrino
direction to about $25^{\circ}$.


\subsection{Double-bang and lollipop events}

Double-bang and lollipop events are signatures unique to tau
neutrinos, made possible by the fact that tau leptons decay before
they lose a significant fraction of their energy~\cite{propagation}.
Double-bang events~\cite{LP,athar2,double3} consist of a hadronic
shower initiated by a charged-current interaction of the $\nu_\tau$
followed by a second energetic shower (hadronic or electromagnetic)
from the decay of the resulting tau lepton.  Lollipop events consist
of the second of the two double-bang showers along with the
reconstructed tau lepton track (the first bang may be detected or
not).  Inverted lollipops, consisting of the first of the two
double-bang showers along with the tau lepton track, are not as useful
as they will often be confused with a hadronic shower in which a $\sim
100$ GeV muon is produced (because of the higher lepton mass, tau
tracks suffer much less catastrophic bremsstrahlung than muons at the
same energy).  We do not consider inverted lollipops for this reason.

The range of a tau lepton is bounded from above by its lifetime in the
lab frame.  A tau lepton of energy $E_\tau$ has a mean lifetime given by
\begin{eqnarray}
R_{\tau}(E_{\nu_\tau},y)
=\frac{E_\tau}{m_\tau}c\tau_{\tau}=
\frac{(1-y)E_{\nu_\tau}}{m_\tau}c\tau_{\tau},
\end{eqnarray}
where $m_\tau$ and $\tau_\tau$ are the mass and rest-frame lifetime.

Following Ref.~\cite{LP,athar2,double3}, the conditions which must be
fulfilled for the detection of a double-bang event are:
\begin{itemize} 
        \item The tau neutrino must interact via the charged-current,
        producing a hadronic shower of sufficient energy to trigger
        the detector ($\sim$1 TeV) inside of the detector volume.

        \item The tau lepton produced in the interaction must decay
        inside the detector volume, producing an electromagnetic or
        hadronic shower of sufficient energy to trigger the detector.

        \item The tau lepton must travel far enough such that the
        two showers are sufficiently separated to be distinguished
        from each other.
\end{itemize}
At the energies required for the third condition to be satisfied, both
the showers will be energetic enough to easily fullfil the threshold
requirements.  The probability for a double-bang event, per incident
tau neutrino, is
\begin{eqnarray}
\label{db}
P_{\rm db}(E_{\nu_\tau}) & \simeq &
\rho N_A \int^1_0 dy \frac{d\sigma}{dy}
\int_{x_{\rm min}}^L \!\! dx \frac{(L-x)}{R_\tau} e^{-x/R_\tau} \nonumber \\
& \simeq & \rho N_A \int^1_0 dy \frac{d\sigma}{dy}
\left[ \left( L-x_{\rm min} -R_\tau \right) e^{-x_{\rm min}/R_\tau}\right.
\nonumber\\
& & + \left. R_\tau e^{-L/R_\tau} \right] \,.
\end{eqnarray}
The track integration from $x_{\rm min}$ to $L$ includes tau lengths
that are larger than $x_{\rm min}$ (for shower separation) and smaller
than $L$ (so that both showers are contained in the detector.)  The
exponential in the integral samples over the decay length
distribution.

Conditions which must be fulfilled for the detection of a lollipop event are:
\begin{itemize} 
        \item The shower produced by the decay of the tau lepton
        must occur within the detector volume and be of sufficient
        energy to trigger the detector ($\sim 1 \rm{\ TeV}$).  This may
        be a hadronic or electromagnetic shower.

        \item The track of the tau lepton must be long enough (within
        the detector) to be reconstructed and separable from the
        shower.  For photomultiplier tube spacing of $\sim 125$
        meters (IceCube's horizontal spacing is 125 meters, their
        vertical spacing is a significantly smaller 17 meters),
        reasonable values for the minimum tau range, $x_{\rm{min}}$,
        are $200 - 400$ meters.
\end{itemize}
The probability for a lollipop event, per incident tau neutrino, is
\begin{equation}
\label{lp}
P_{\rm lollipop}(E_{\nu_\tau}) \simeq
\rho N_A (L - x_{\rm min})
\int^1_0 dy \frac{d\sigma}{dy} e^{-x_{\rm min}/R_\tau}\,.
\end{equation}
Note that the energy threshold for both double-bangs and lollipops,
resulting from the requirement $R_\tau \gtrsim x_{\rm min }$, is given
by $E^{\rm thr}_{\nu_\tau}\sim 5\,(x_{\rm min }/250 {\rm\ m}) {\rm\
PeV}$.  For double-bang events, the energy threshold for the first
shower, $E_{\rm sh}\sim y\,E_{\nu_{\tau}}\gtrsim{\rm TeV}$ puts a
lower limit on the $y$-integration, but negligibly so for PeV scale
neutrino energies.

\begin{figure}[t]
\includegraphics[height=3.25in,angle=90]{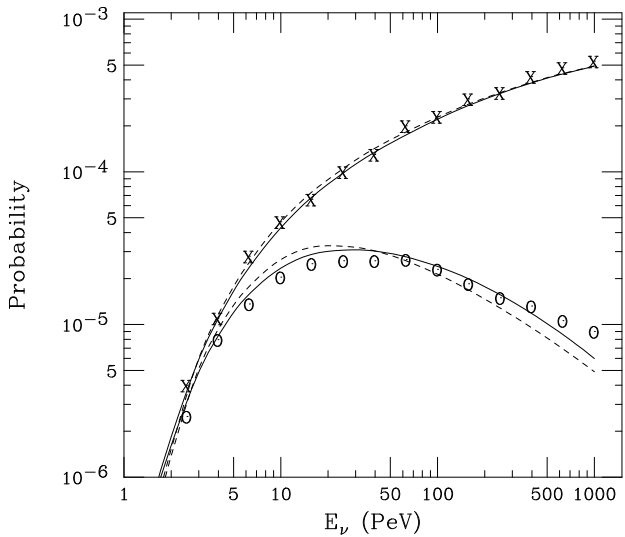}
\caption{\label{fig:tau} Probability of observing a lollipop (upper)
or double-bang (lower) event in IceCube per incident tau neutrino.
The points (X's and O's) represent Monte Carlo results while the solid
lines represents the analytic expressions of Eqs.~(\ref{db}) and
(\ref{lp}) and the dashed lines represents the approximations of
Eqs.~(\ref{dbapprox}) and (\ref{lpapprox}).  We have set $L = 1$ km
and $x_{\rm{min}} = 250$ m.  Above about 20 PeV, the tau track length
becomes long enough that it may exceed the thickness of the ice for
near-downgoing events, introducing zenith-angle dependence.}
\end{figure}

The expressions in Eqs.~(\ref{db}) and (\ref{lp}) can be further
simplified with the approximation
\begin{equation}
\label{yapprox}
\frac{d\sigma}{dy} \simeq \sigma \delta(y - \langle y\rangle),
\end{equation}
where $\langle y\rangle \simeq 0.25$ at PeV scale energies and
$\sigma$ is the total cross section.  Thus,
\begin{eqnarray}
\label{dbapprox}
P_{\rm db}(E_{\nu_\tau}) & \simeq &
\rho N_A \sigma
\left[\left( L-x_{\rm min} - R_\tau \right) e^{-x_{\rm min}/R_\tau}\right.
\nonumber \\
& & \hspace{1cm} + \left. R_\tau e^{-L/R_\tau}
\right]_{y=\langle y\rangle}\,, \\
\label{lpapprox}
P_{\rm lollipop}(E_{\nu_\tau}) &\simeq& \rho N_A \sigma 
(L - x_{\rm min})
\left[e^{-x_{\rm min}/R_\tau} \right]_{y=\langle y\rangle}\,.
\end{eqnarray}

Fig.~\ref{fig:tau} shows the lollipop and double-bang detection
probabilities.  Note that near threshold, the majority of lollipop
events are also double-bang events.  At higher energies, the size of
the detector excludes many double-bang events but not lollipops.

The analytic expressions in Eqs.~(\ref{db}) to (\ref{lpapprox}) are
original to this work.  To check their validity, we constructed a
Monte Carlo to calculate the probabilities of observing double-bang or
lollipop events in IceCube.  This simulation took into account the
distribution of $y$ values in charged-current interactions as well as
the variation in decay lengths.  In common with the analytic results,
the Monte Carlo treats the detector one dimensionally.  We note that
this is a conservative approximation as taus that travel diagonally
across the detector could have bangs separated by distances larger up
to a factor $\sqrt{3}$.  The analytic expressions and the Monte Carlo
calculation shown in Fig.~\ref{fig:tau} are in good agreement, a
considerable improvement over the approximations given in
Ref.~\cite{athar2,double3}.  In particular, the important threshold
region is characterized very well.  The remaining minor discrepancies
we attribute to the low statistics of the Monte Carlo; our formulas
should be quite accurate.  In addition, the approximation in
Eq.~(\ref{yapprox}) is seen to yield quite accurate results (which
indicates that the variation of cross section with $y$ is relatively
unimportant in these calculations).  Thus we have good confidence that
Eqs.~(\ref{db}) to (\ref{lpapprox}) may join with Eqs.~(\ref{muprob})
to (\ref{showerprob}) to complete the set of probabilities of event
topologies.


\subsection{Determining the Muon/Shower Ratio}

The spectrum of shower events is more difficult to infer from data
than is the muon spectrum due to lower statistics, but it may be
possible \cite{brazil}.  Relative disadvantages are the considerably
higher shower energy threshold~\cite{IceCube} and the fact that muon
event rates benefit from long muon ranges.  However, in contrast to
the muon signals, shower events are produced by all flavors of
neutrinos, and shower energies more faithfully represent the neutrino
energy than do muon tracks.  We do not attempt to relate the observed
shower spectrum to the spectrum of incident neutrinos.  Rather, we
assume that, due to oscillations, the neutrino spectrum shape is
independent of flavor; i.e., the neutrino spectrum inferred from a
measurement of the muon spectral shape is universal.  We may use this
universal spectrum to produce shower events and muon tracks, and then
compare the total number of shower events to the total number of muon
events to obtain the flavor ratios.  Recall that $\nu_\mu - \nu_\tau$
symmetry means 
$\phimu:\phitau=1:1$, and so two independent
observables are sufficient to determine all three flavor ratios.

For a neutrino spectrum of $E^2_{\nu_{\mu}}
dN_{\nu_{\mu}}/dE_{\nu_{\mu}} =
10^{-7} \rm{\ GeV} \rm{\ cm}^{-2} \rm{\ s}^{-1}$ for one year,
we expect, on average, 323 muon events (186 of which have contained
vertices). Given an equal number of each neutrino flavor, we predict
only 36 shower events. At the 68\% confidence level, this allows for a
measurement of $N_{\rm{muons}}/N_{\rm{showers}}=9.0^{+1.6}_{-1.9}$
for horizontal sources, about a 20\% uncertainty. For upgoing sources,
a measurement of
$N_{\rm{muons}}/N_{\rm{showers}}=8.5^{+1.5}_{-1.8}$ results
similarly.  The difference comes from the fact that
upgoing neutrinos are absorbed at high energies, where muon events are
more likely to occur, thus slightly lowering the muon to shower ratio
for upgoing events relative to horizontal events.  These results use
the natural energy thresholds of the detector, $\sim 100$ GeV for
muons and $\sim 1$ TeV for showers.


\subsection{Signatures Unique To Tau Neutrinos}

Given the flux we have considered in this paper,
$E^2_{\nu_{\mu}} dN_{\nu_{\mu}}/dE_{\nu_{\mu}} =
10^{-7} \rm{\ GeV} \rm{\ cm}^{-2} \rm{\ s}^{-1}$ for one year,
we predict on the order of a 50\% chance of observing a lollipop
event, and similarly for a double-bang event.  Thus IceCube is
unlikely to provide a stringent probe of the tau neutrino flux.  The
lack of observed double-bangs or lollipops in IceCube would not reveal
much, though the positive identification of even a single such event by
IceCube would indicate the important existence of a tau neutrino flux
on the order of the flux we consider here.

Even larger detectors are needed to exploit the double-bang and
lollipop features.  With several of these events, flavor ratios could
be easily reconstructed, and the $\nu_\mu - \nu_\tau$ symmetry tested.
Since even larger detectors are even farther into the future, we do
not consider the double-bang and lollipop signatures further in this
work.


\section{Inferring Neutrino Flavor Ratios}

Although the neutrino flavor ratios are not directly accessible at
neutrino telescopes, the indirect flavor information collected from
such experiments, i.e., the ratios of muon, shower and tau-unique
events, can be very useful in inferring flavor information.

\begin{figure}
\includegraphics[height=3.25in,angle=90]{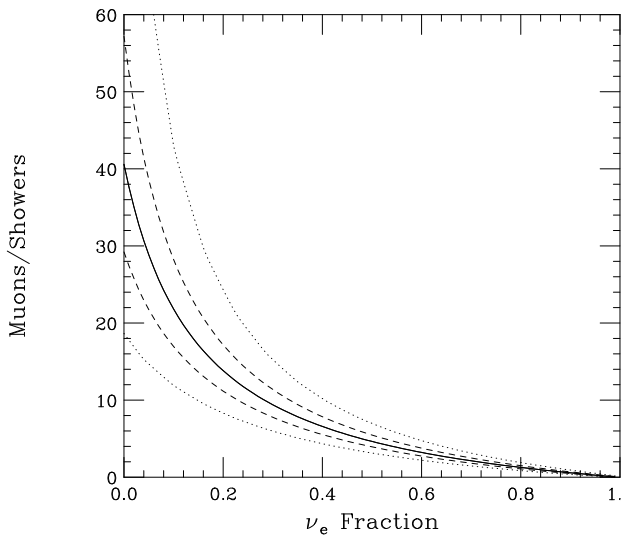}
\caption{\label{fig:ratio} The relationship between the muon-to-shower
ratio and the $\nu_e$ fraction, assuming $\nu_\mu - \nu_\tau$ symmetry 
and an $E^{-2}$ power law spectrum. The muon energy threshold is 100 GeV,
and the shower threshold is 1 TeV.  Horizontal sources are assumed.
The solid line is the central predicted value.  The dashed lines
represent the 68\% confidence interval for a spectrum of
$E^2_{\nu_{\mu}} dN_{\nu_{\mu}}/dE_{\nu_{\mu}} =
10^{-7} \rm{\ GeV} \rm{\ cm}^{-2} \rm{\ s}^{-1}$ for one year.
The dotted lines represent the 68\% confidence interval for a spectrum
five times smaller.}
\end{figure}

In the previous section, we showed that for a neutrino spectrum of
$E^2_{\nu_{\mu}} dN_{\nu_{\mu}}/dE_{\nu_{\mu}} =
10^{-7} \rm{\ GeV} \rm{\ cm}^{-2} \rm{\ s}^{-1}$ for one year,
IceCube could determine the ratio of the muon events to shower events
with uncertainties on the order of 20\%. Given $\nu_\mu - \nu_\tau$
symmetry, this ratio can be used to deduce the ratio of electron
neutrinos to either muon or tau neutrinos. Clearly, precision
measurements of these quantities are unlikely to be determined in this
fashion. Fortunately, in some interesting theoretical scenarios, the
predictions for deviations from a 1:1:1 flavor ratio are so extreme as
to not require greater precision.  The neutrino decay
model~\cite{decay} provides a splendid example of possibly large
flavor deviations.

\begin{figure}[t]
\includegraphics[height=3.25in,angle=90]{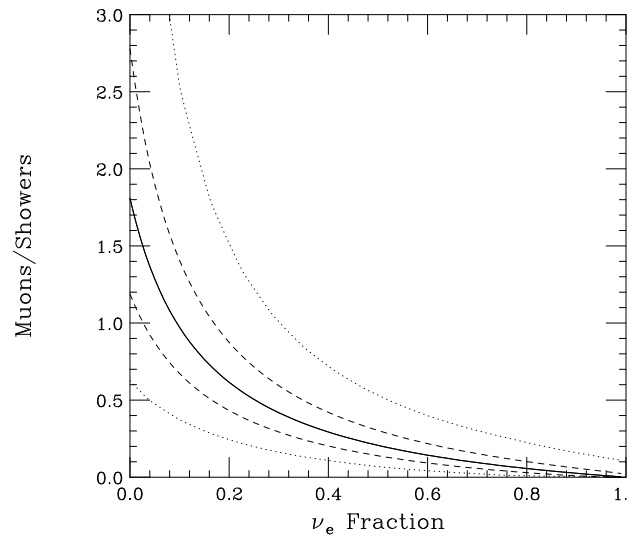}
\caption{\label{fig:ratio2} Same as Fig.\ \ref{fig:ratio} except that
a 1 TeV muon energy threshold has been imposed to reduce the effect of
the uncertainty in the muon spectrum.}
\end{figure}

In Fig.~\ref{fig:ratio}, we show the relationship between the
muon-to-shower ratio and $\nu_e$ fraction, assuming
$\nu_\mu - \nu_\tau$ symmetry and an $E^{-2}$ power law
spectrum. We define the $\nu_e$ fraction as the fraction of neutrinos
with electron flavor (1 for all electron neutrinos, 0 for no electron
neutrinos). The solid line is the central predicted value. The dashed
lines represent the 68\% confidence interval for our chosen spectrum.
The dotted lines represent the 68\% confidence interval for an
exposure five times smaller. Note that the $\nu_e$ fraction of
1/3, expected from known particle physics, can be measured to a range
of 0.26-0.37 or 0.18-0.44 for the two fluxes shown in
Fig.~\ref{fig:ratio}, respectively. These measurements are sufficient
to test interesting speculations such as neutrino decay.

If the neutrino spectrum is not well measured, then the relationship
between the flavor ratios and the muon/shower ratio will be
altered. In the case of poor muon spectrum resolution, some steps
could be taken to reduce the related uncertainties. For example, the
number of shower events could be compared to muon events above a
threshold near 1 TeV, rather than the experimental threshold near 100
GeV. By doing this, the effect of the spectral slope is reduced, as
shower events and muon events only from neutrinos with similar
energies are compared.  In Fig.~\ref{fig:ratio2}, we show results
analogous to Fig.~\ref{fig:ratio}, but with a muon energy threshold of
1 TeV imposed (Dutta, Reno, and Sarcevic~\cite{renodutta}
considered a showers/muons ratio to test three-flavor active neutrino
oscillations against no oscillations, assuming standard flavor ratios
at production; our results are in reasonable agreement at the one
point of common consideration, a $\nu_e$ fraction of 1/3).  With this
choice of threshold, a measurement of $\alpha=2.0\pm0.2$ corresponds
to an uncertainty on the order of only 20\% in the predicted
muon/shower ratio.  A $\nu_e$ fraction of 1/3 can be measured to a
range of 0.22-0.42 or 0.09-0.61 for the two fluxes shown in
Fig.~\ref{fig:ratio2}, without using any information from muons below
1 TeV.  With Fig.~\ref{fig:ratio} as our guide, we summarize in
Table~\ref{table:one} the muon-to-shower ratio expected for each of
various astrophysical neutrino models.


\begin{table}
\caption{\label{table:one} Summary of the muons/showers ratios
expected for selected scenarios.  The decay scenarios with normal and
inverted neutrino mass hierarchies are taken from Ref.~\cite{decay}.}
\begin{tabular}{c|c|c|c|c}
\hline\hline
ratios at & decays & ratios at & $\nue$ fraction & muons/showers\\ 
source    &        & Earth     &                 & (Fig.\ 7 central value)\\
\hline
$1:2:0$   & none     & $1:1:1$ & 0.33            & 9    \\ \hline
          & Normal   & $6:1:1$ & 0.75            & 1.5  \\ \hline
          & Inverted & $0:1:1$ & 0               & 40   \\ \hline
$0:1:0$   & none     & $1:2:2$ & 0.2             & 14   \\
\hline\hline
\end{tabular}
\end{table}

\section{Conclusions}

Next generation high energy neutrino telescopes, such as the kilometer
scale experiment IceCube, will be capable of observing muon tracks
produced by muon neutrinos, shower events from all flavors of
neutrinos, and possibly the double-bang and lollipop topologies unique
to tau neutrinos.  Using these features, it will be possible to infer
the flavor ratios of astrophysical neutrino fluxes.  These can be used
to probe properties of the sources~\cite{Barenboim} and to test for
new physics beyond the standard model of neutrino
physics~\cite{decay,Barenboim,pDirac}.

Muon events, being the most numerous in under-ice/water detectors, are
the most useful for measuring the astrophysical neutrino spectrum.
We demonstrate that this can be accomplished with acceptable accuracy,
given a sufficiently large but realistic neutrino flux.  Knowledge of
the spectral shape then allows us to make comparisons of the total
number of event types (muons, showers, or tau-unique events) to infer
the neutrino flavor ratios.

Assuming $\nu_\mu - \nu_\tau$ symmetry as indicated by oscillation
data, the ratio of showers to muons provides sufficient information to
determine all three neutrino flavor ratios. Large deviations from the
flavor ratios predicted by oscillations (1:1:1) will likely be
observable.  Furthermore, tau-unique events (double-bangs and
lollipops) provide another tool with which to address flavor
identification; however, unless there is a larger than expected tau
neutrino flux, such events will probably be too small in number in
kilometer-scale detectors to provide useful information beyond the
existence of astrophysical tau neutrinos.


\smallskip

\begin{acknowledgments}
We thank Steve Barwick, 
Doug Cowen, Peter Gorham, Francis Halzen, and John Learned
for valuable discussions.  J.F.B. and N.F.B. were supported by
Fermilab (operated by URA under DOE contract DE-AC02-76CH03000) and by
NASA grant NAG5-10842.  D.H was supported by DOE grant
DE-FG02-95ER40896 and the Wisconsin Alumni Research Foundation,
S.P. by DOE grant DE-FG03-94ER40833 and T.J.W. by DOE grant
DE-FG05-85ER40226.  NFB, SP, and TJW thank the Kavli Institute for
Theoretical Physics at the University of California, Santa Barbara,
for support and hospitality.
\end{acknowledgments}



\onecolumngrid
\newpage

\begin{center}
{\large\bf Erratum: Measuring Flavor Ratios of High-Energy
Astrophysical Neutrinos}\\
{\large\bf [Phys. Rev. D 68, 093005 (2003)]}

\bigskip

{\bf John F. Beacom}\\
{\tt beacom@mps.ohio-state.edu}
\medskip

{\bf Nicole F. Bell}\\
{\tt nfb@caltech.edu}
\medskip

{\bf Dan Hooper}\\
{\tt hooper@astro.ox.ac.uk}
\medskip

{\bf Sandip Pakvasa}\\
{\tt pakvasa@phys.hawaii.edu}
\medskip

{\bf Thomas J. Weiler}\\
{\tt tom.weiler@vanderbilt.edu}

\bigskip
(Erratum Date: 15 June 2005)

\end{center}


Due to a programming error, some of the results presented in our
published paper~[1] are incorrect.  Here we revise
Figs.~1--4,7,8 and some of the associated discussion, including the
final column of Table I.  The conclusions of the paper are mostly
unchanged, and in some cases improved, despite large changes in some
of the intermediate results.

Figures 1--4 show the observable muon spectra for the IceCube
experiment, resulting from the given neutrino spectra; the new curves
all fall less steeply with energy than before.  Above 1 TeV, the new
curves are higher by up to an order of magnitude.  Below 1 TeV, the
new curves are slightly lower for ``All Muons'' (meaning a trigger on
all muon tracks), and an order of magnitude lower for ``Muons
(Contained Vertex)''.  Thus, the statistical uncertainties will be
improved when muons above 1 TeV are used, which is preferred, since
this is well matched to the 1 TeV threshold required for resolving
showering events.  (Previously, we had argued that using a muon
threshold of 100 GeV was statistically beneficial, at the cost of some
model dependence; this argument is no longer necessary.)  
As before, contained vertex muons offer a more robust neutrino energy
estimate than do the throughgoing muons that dominate the all-muons
sample, although the statistics of the all-muons sample is better.
The rate of the contained events is still about 1/10 of the all-muons
sample, but now the absolute statistics of both are more favorable.

Figures 7 and 8 reveal graphically the most important results of our
paper.  They show the expected ratios of muon events to
showering events (all events other than muon-track events), as a
function of the $\nu_e$ fraction of the neutrino flux.  The
IceCube experimental efficiencies are folded into these figures.
Thus, these figures provide the map from the incident flavor fraction
onto a direct experimental observable, and vice versa.  (We assume
$\nu_\mu\leftrightarrow\nu_\tau$ interchange symmetry, which is
supported by inferences from neutrino oscillation results.)  The
qualitative message of these figures is the same as before: an IceCube
measurement of the muon/shower ratio can determine the flavor content
of the incident neutrino beam with sufficient precision to study
neutrino physics and astrophysical dynamics.

While the normalizations of the new curves are quite different, the
statistics are increased, so that the final conclusions are quite
similar.  For example, consider an incident $\nu_e$~fraction of $1/3$,
as expected from pion decay, and note the predicted muon/shower ratio.
Using the dashed and dotted lines, corresponding to the uncertainties,
note also the range of different incident $\nu_e$ fractions that would
give the same muon/shower ratios.  In this standard case, the $\nu_e$
fraction would be deduced to be in the ranges 0.25--0.40 and
0.14--0.50 for the two fluxes in the new Fig.~7 (100 GeV muon
threshold), and 0.24--0.41 and 0.12--0.52 for the two fluxes of the
new Fig.~8 (1 TeV muon threshold).  The ranges quoted in the published
paper were similar: 0.26--0.37 and 0.18--0.44 for the lower muon
threshold, and 0.22--0.42 and 0.09--0.61 for the higher muon
threshold.  The muon/shower ratios in the final column of Table~I
should be changed from the original values of (9, 1.5, 40, 14) for
$\nu_e$-fraction (0.33, 0.75, 0, 0.2), to (3, 1.0, 5, 3.7) for the
same $\nu_e$-fractions.  Thus, while the normalizations of the curves
have changed, the precision with which the flavor ratios can be
measured has not.

In summary, with the corrections to our paper, the measurements of
neutrino spectra and flavor ratios are made easier, especially above 1
TeV.  The other results of the paper, including the detection
probabilities for different neutrino flavors, are unchanged.

We are grateful to Shin'ichiro Ando for pointing out that our results
were in error, and for subsequent discussions.

\begin{figure}
\includegraphics[width=0.38\columnwidth,angle=90]{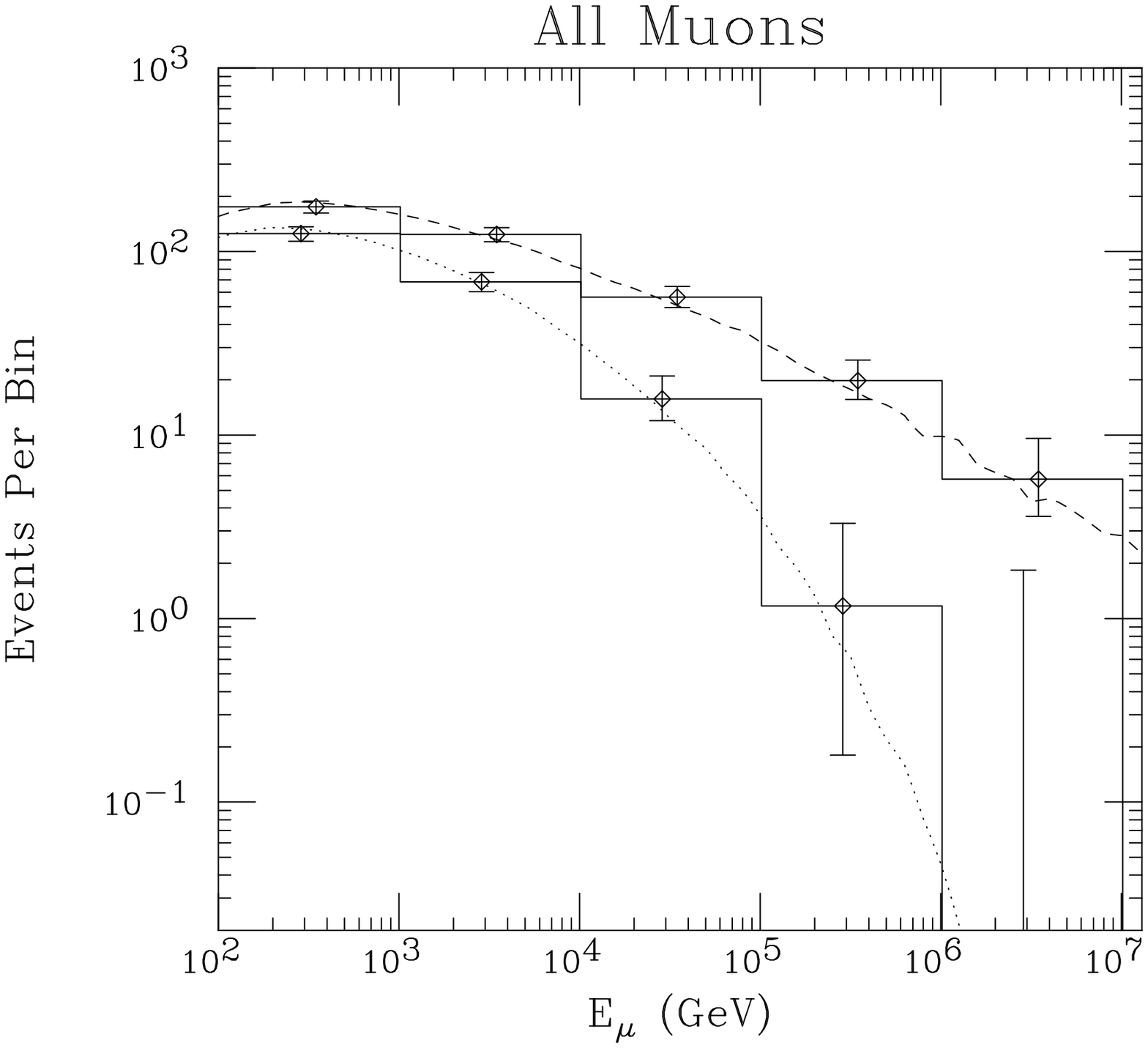}
\hspace{0.5cm}
\includegraphics[width=0.38\columnwidth,angle=90]{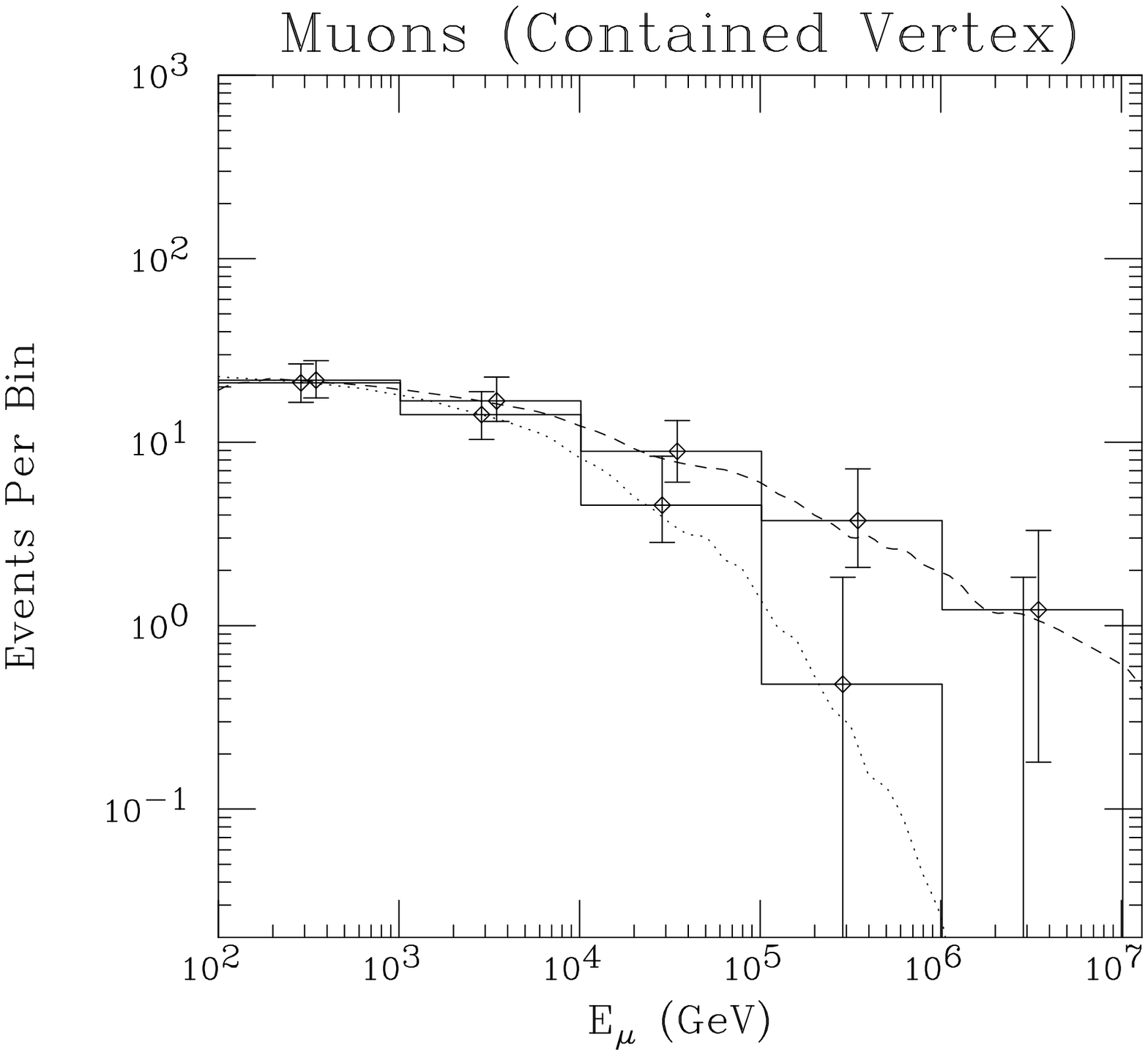}
\caption{\label{fig:muons} The replacements for Figs. 1 and 2.}
\end{figure}

\begin{figure}
\includegraphics[width=0.38\columnwidth,angle=90]{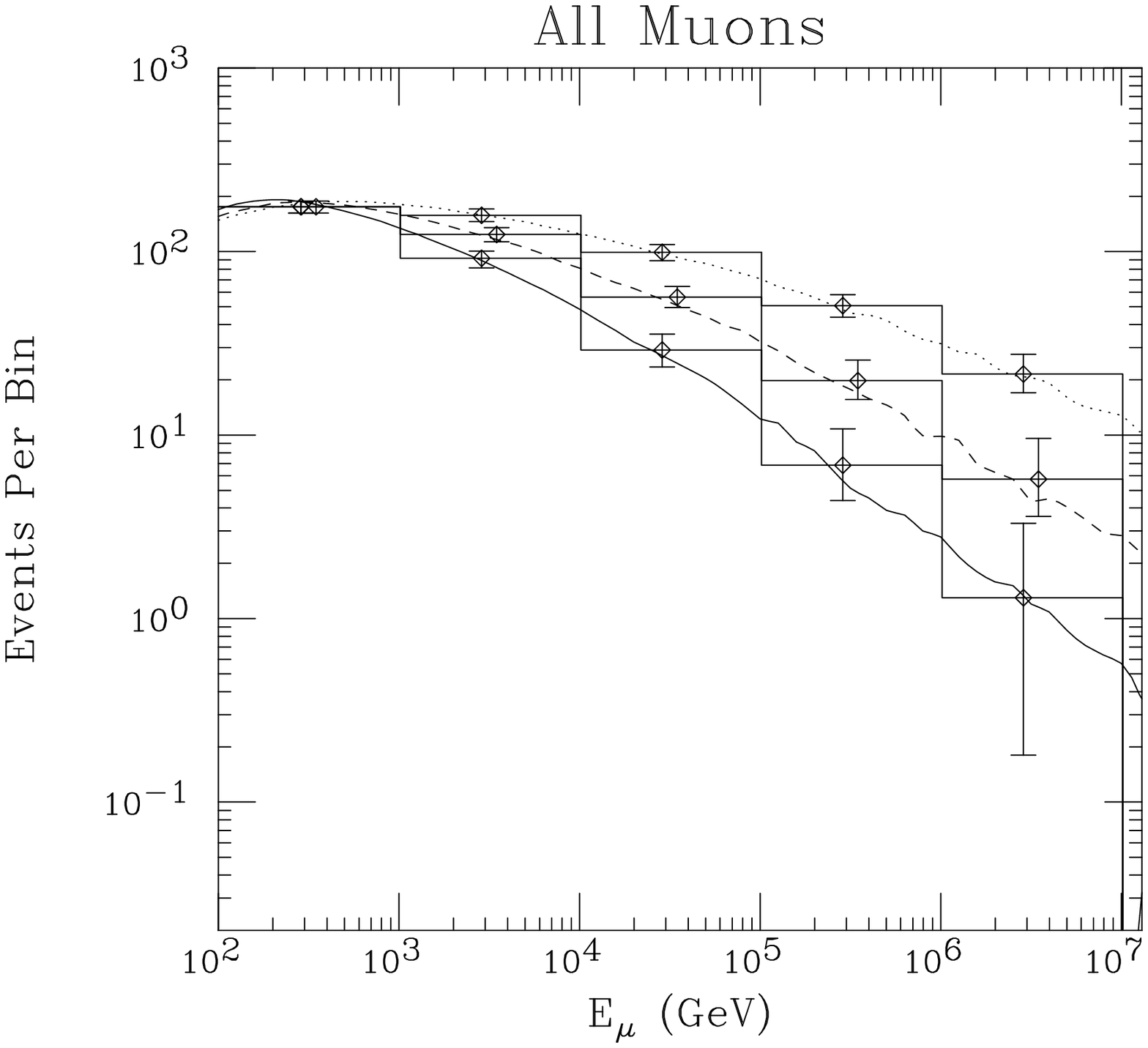}
\hspace{0.5cm}
\includegraphics[width=0.38\columnwidth,angle=90]{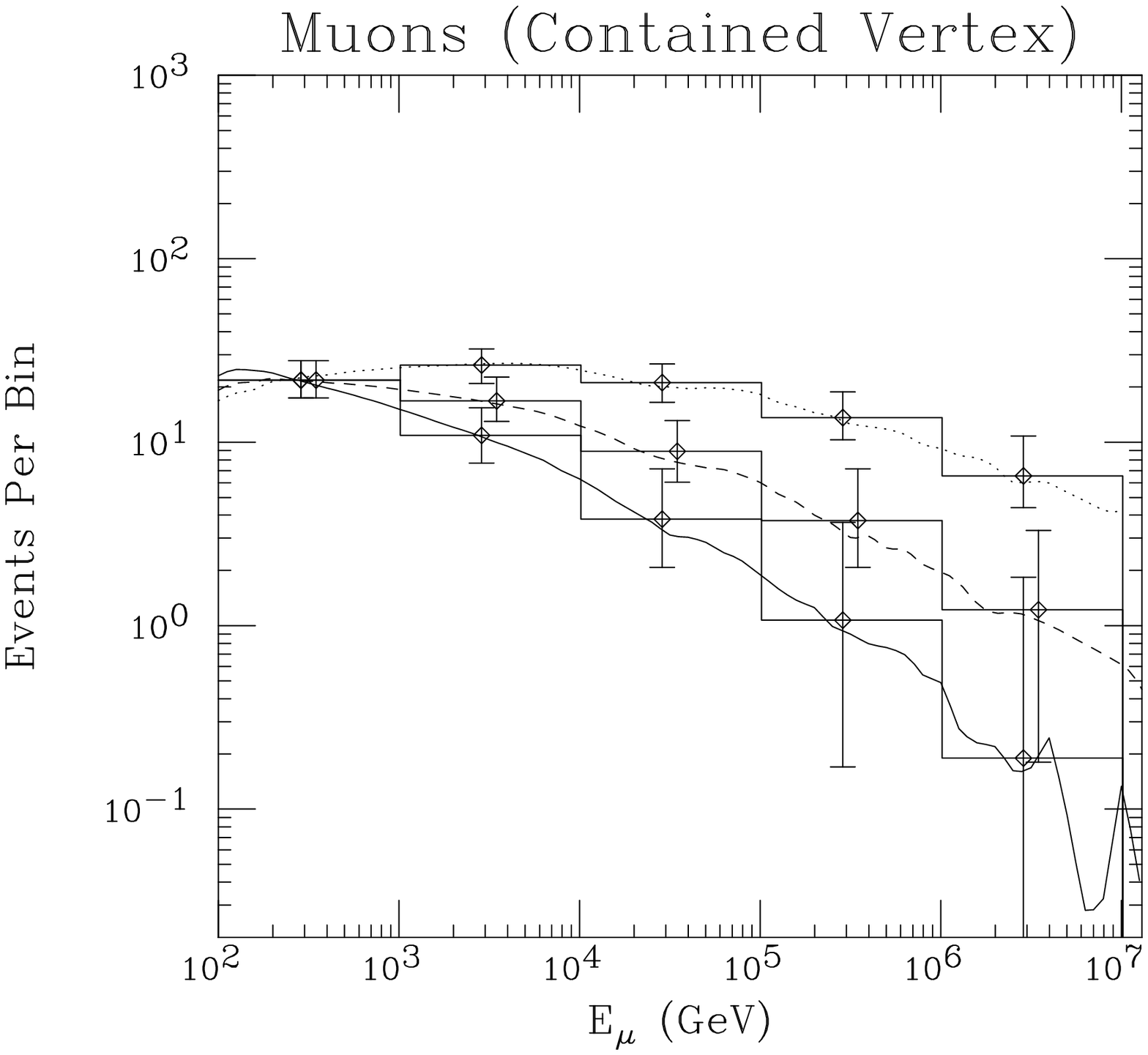}
\caption{\label{fig:spectra} The replacements for Figs. 3 and 4.}
\end{figure}

\begin{figure}
\includegraphics[width=0.35\columnwidth,angle=90]{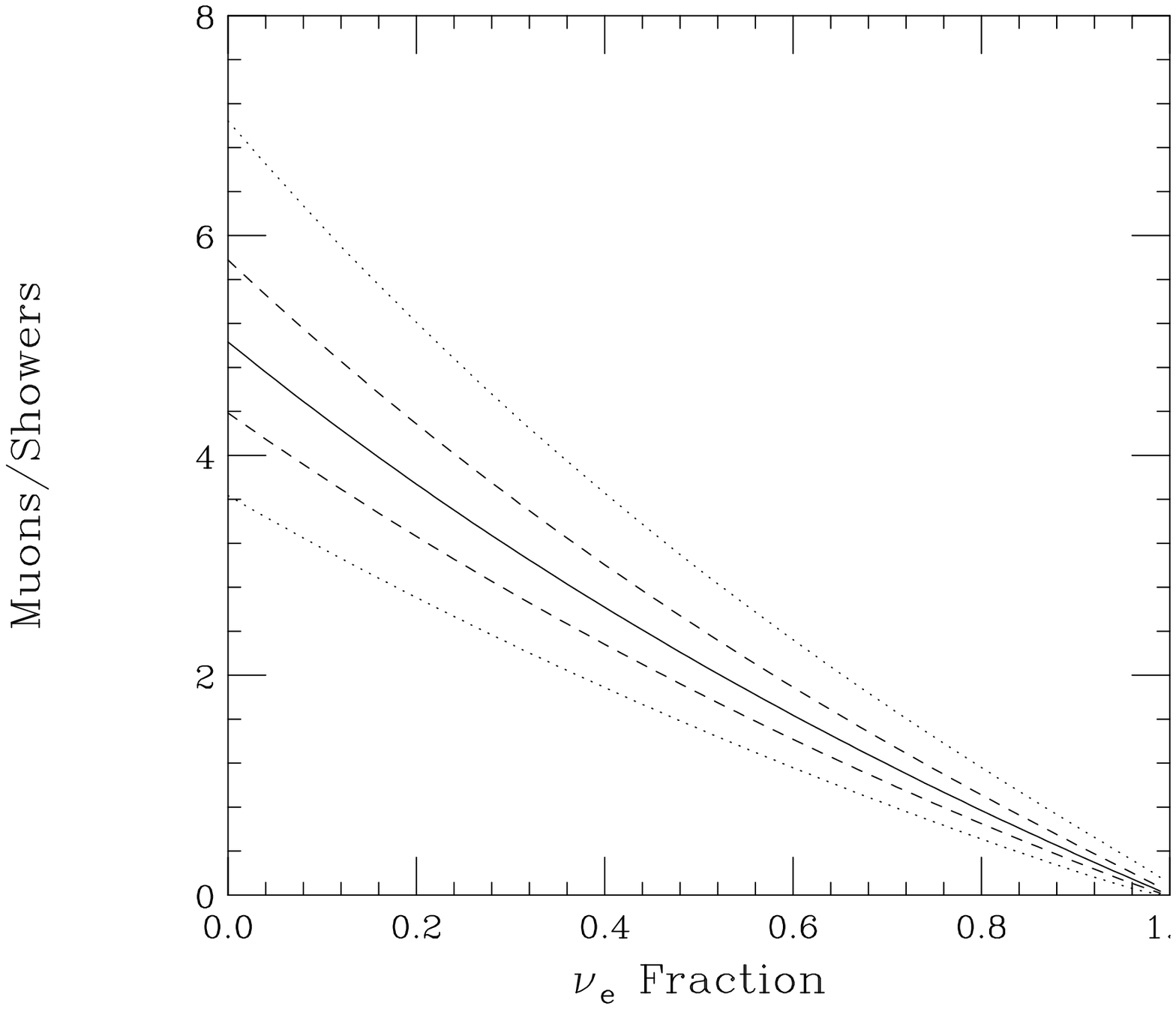}
\hspace{0.5cm}
\includegraphics[width=0.35\columnwidth,angle=90]{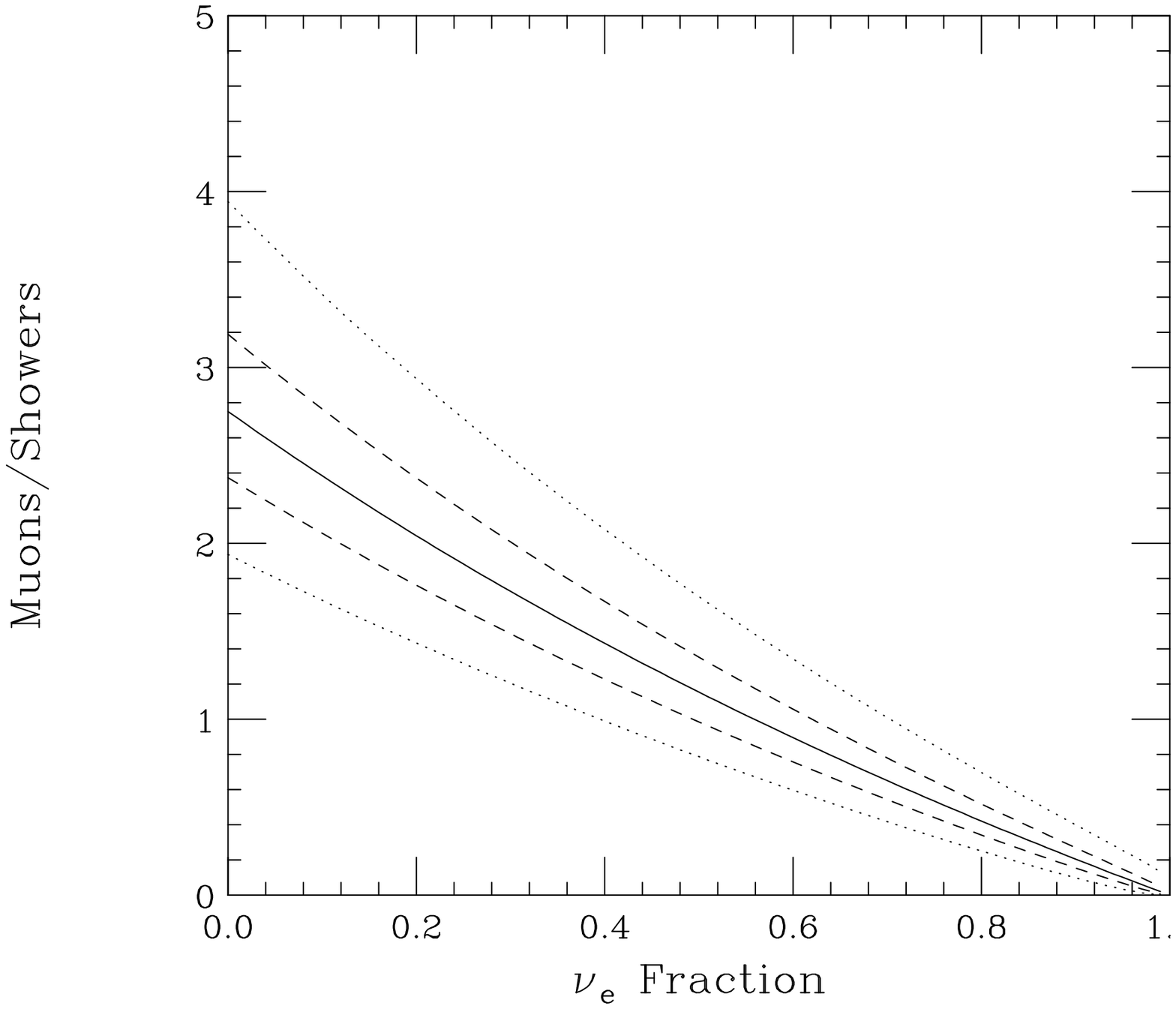}
\caption{\label{fig:ratios} The replacements for Figs. 7 and 8.}
\end{figure}

\bigskip

[1] J.~F.~Beacom, N.~F.~Bell, D.~Hooper, S.~Pakvasa and T.~J.~Weiler,
``Measuring flavor ratios of high-energy astrophysical neutrinos,''
Phys.\ Rev.\ D {\bf 68}, 093005 (2003) [arXiv:hep-ph/0307025].

\end{document}